# Machine Learning for Anomaly Detection and Categorization in Multi-cloud Environments

Tara Salman
Washington University in St. Louis,
St. Louis, USA
tara.salman@wustl.edu

Deval Bhamare
Qatar University,
Doha, Qatar
devalb@qu.edu.qa

Aiman Erbad
Qatar University,
Doha, Qatar
aerbad@qu.edu.qa

Raj Jain
Washington University in St. Louis,
St. Louis, USA
jain@wustl.edu

Mohammed Samaka
Qatar University,
Doha, Qatar
Samaka.m@qu.edu.qa

*Abstract*— Cloud computing has been widely adopted by application service providers (ASPs) and enterprises to reduce both capital expenditures (CAPEX) and operational expenditures (OPEX). Applications and services previously running on private data centers are now being migrated to private or public clouds. Since most of the ASPs and enterprises have globally distributed user bases, their services need to be distributed across multiple clouds, spread across the globe which can achieve better performance in terms of latency, scalability and load balancing. The shift has eventually led the research community to study multi-cloud environments. However, the widespread acceptance of such environments has been hampered by major security concerns. Firewalls and traditional rule-based security protection techniques are not sufficient to protect user-data in multi-cloud scenarios. Recently, advances in machine learning techniques have attracted the attention of the research community to build intrusion detection systems (IDS) that can detect anomalies in the network traffic. Most of the research works, however, do not differentiate among different types of attacks. This is, in fact, necessary for appropriate countermeasures and defense against attacks. In this paper, we investigate both detecting and categorizing anomalies rather than just detecting, which is a common trend in the contemporary research works. We have used a popular publicly available dataset to build and test learning models for both detection and categorization of different attacks. To be precise, we have used two supervised machine learning techniques, namely linear regression (LR) and random forest (RF). We show that even if detection is perfect, categorization can be less accurate due to similarities between attacks. Our results demonstrate more than 99% detection accuracy and categorization accuracy of 93.6%, with the inability to categorize some attacks. Further, we argue that such categorization can be applied to multi-cloud environments using the same machine learning techniques.

## I. INTRODUCTION

Since its inception, cloud computing has attracted a lot of interest from both, industry and academia. It is widely adopted by organizations and enterprises as a cost effective and a scalable solution that reduces both capital expenditures (CAPEX) and operational expenditures (OPEX). Cloud computing allows application service providers (ASPs) to deploy their applications, store, and process the data without physically hosting servers [1]. Nowadays, clouds hosting services and data are geographically distributed to be closer in proximity to the end-users. Such a networking paradigm is called as a multi-cloud environment. Thus, in a multi-cloud environment, multiple cloud service providers (CSPs), ASPs, and network service providers (NSPs) collaborate to offer various services to their end-users. Such multi-cloud approach is getting popular within organizations, clients, and service providers [2]. Despite this fact, data security is a major concern for the end-users in multi-cloud environments. Protecting such environments against attacks and intrusions is a major concern in both research and industry [3].

Firewalls and other rule-based security approaches have been used extensively to provide protection against attacks in the data centers and contemporary networks. However, large distributed multi-cloud environments would require a significantly large number of complicated rules to be configured, which could be costly, time-consuming and error prone [4]. Furthermore, the current advances in the computing technology have aided attackers in the escalation of attacks as well, for example, the evolution of Denial of Service (DoS) attacks to Distributed DoS (DDoS) attacks which are rarely detected by traditional firewalls [5]. Thus, the use of firewalls alone is not sufficient to provide full system security in multi-cloud scenarios.

Recently, advances in machine learning techniques have proven their efficiency in many applications, including intrusion detection systems (IDS). Learning-based approaches may prove useful for security applications as such models could be trained to counter a large amount of evolving and complex data using comprehensive datasets. Such learning models can be incorporated with firewalls to improve their efficiency. A well trained model with comprehensive attack types would improve anomaly detection efficiency significantly with a reasonable cost and complexity [6]. A significant amount of research has been already made towards the development of IDS models in recent years. Several such examples are summarized in surveys such as [7] [8]. However, such works suffer from two burdens: categorization among different attack types and their applicability to multi-cloud environments. Most of the researchers in this domain have been tackling anomaly detection problem without paying much attention to categorizing the attacks. We argue that such categorization is critical to identify system vulnerabilities and impose appropriate countermeasures and defense against different types of attacks.

Multi-cloud environments are new and have not been studied in the literature from a security perspective using machine learning techniques. The reason could be non-availability of the comprehensive training datasets that include samples of recent attacks in such environments. Datasets for security are rarely available, mostly due to privacy issues. Common publicly available datasets mostly include a single type of attack only [9]. Furthermore, research works that use datasets with multiple attacks mostly focus on anomaly detection rather than attack classification or categorization.

In this paper, we investigate the feasibility of applying machine learning techniques to distinguish between different attacks rather than just detecting anomalous traffic. In other words, we take anomaly detection systems a step further by enabling them to identify the specific types of attacks. We have considered a new and publicly available dataset given by UNSW [10, 11]. We use supervised learning to build anomaly detection models and demonstrate their detection efficiency. Later on, we build learning models to distinguish between different types of attacks, and we refer to that as attack categorization. To achieve categorization, we apply two strategies: a classical strategy and our own proposed strategy. We refer to the classical one as *single-type attack categorization* while we propose a *step-wise attack categorization*. We demonstrate more than 99% anomaly detection accuracy and 93.6% categorization accuracy. Such results have better categorization accuracy than *single-type detection*, however, similarities between some of UNSW attacks resulted in high misclassification error for those attacks. We argue that more distinguishable data are needed to classify those attacks accurately.

The rest of the paper is organized as follows: Section II gives a brief background of state of the art, publicly available datasets, and our contributions. In Section III, we present the anomaly detection models with the feature selection algorithm and the supervised machine learning models. Section IV presents both single-type and step-wise attack categorization strategies along with results and comparisons. Finally, Section V concludes the paper.

## II. BACKGROUND

In this section, we present state of the art in the domain of machine learning techniques for network and cloud security. In addition, we discuss the important datasets available for the training of the machine learning models. Furthermore, we highlight our contribution.

### A. Related Work

A significant amount of work has been performed in the literature for IDS using machine learning techniques. For example, in [12, 13] the authors used support vector machines (SVM) to detect anomalies using KDD dataset [14]. In [15], the authors utilized artificial neural networks to build IDS models for anomaly detection using the same dataset. In [16], the author used cascaded classifiers to detect and classify anomalies in KDD dataset even if such anomalies were unequally distributed. Anomaly detection using decision trees and random forest (RF) has been proposed in [17] [18], respectively. Furthermore, hybrid approaches using two or more machine learning techniques have been proposed in [19]. Results demonstrate that the hybrid approaches achieve better performance than the single models. For more details of machine learning techniques in IDS, readers may refer to the surveys presented in [7], [8], and [20].

The authors in [21] used a four-layered classification approach to detect four types of attacks using KDD dataset. The results demonstrate a small misclassification error and small overall error. In addition, the authors proposed to reduce the number of features in the original dataset which resulted in reduced complexity and better accuracy. However, the authors did not report the misclassification errors resulting from a single type of attack being misclassified as other types. The work in [22] presented such details with the same dataset using supervised, unsupervised and outliers learning techniques. The results showed that some attacks were misclassified, and hence the overall accuracy was lower than the work presented in [21].

KDD dataset has been widely used to build anomaly detection and classification models using machine learning techniques. KDD dataset includes four types of attacks which have completely different traffic behaviors. An approach for classification of attacks using KDD is presented in [21] and it demonstrated a relatively low misclassification error. However, such models may not perform well with current multi-cloud scenarios with evolving and closely related different types of attacks. In addition KDD dataset is becoming old and may not reflect the contemporary real-time traffic patterns and network attacks [23] [24]. Moreover, if a new attack is introduced, it would mostly go undetected and may result in a high error rate as reported in [22].

In addition to KDD, CAIDA [9], UNSW [10, 11], ISOT [25], CDX [26], and ISCX [22], are among other datasets that are publicly available in the literature for the researchers. ISOT and CAIDA cover only single-type of attack, that is botnet and DDoS attacks, respectively. ISCX and CDX datasets do not represent real world traffic as claimed by [27]. Hence, these datasets might not be suitable for classification of different types of attacks using machine learning models. On the contrary, the UNSW dataset, released in 2015, includes nine different types of attacks which are common in the contemporary multi-cloud environments and hence more suitable to be used in the contemporary anomaly detection schemes. Therefore, we distinguish our work from the classification work done in [21] and [22] by including nine attack types, which are highly co-related and are common in the contemporary multi-cloud environments.

### B. Contribution to the Analysis of the UNSW Dataset

We urge the readers to refer to [10] for a detailed description of the UNSW dataset. Key points are summarized below. To build the UNSW dataset, packets were generated using IXIA *PerfectStorm* tool for the realistic modern normal activities and the synthetic contemporary attack behaviors in the network traffic. Then, tcpdump files were collected and used to extract 49 features. Those features were extracted using Argus and Bro network monitoring tools. The collected data were divided into training and testing sets to be used for learning and prediction of attack behaviors, respectively. Statistics about the nine types of UNSW attacks along with normal data are summarized in Table I.

Compared to the KDD dataset, limited research has been conducted on the UNSW [10, 11] dataset. In [10, 11] the authors present accuracy, recall, and precision using UNSW and compare it to KDD performance. We distinguish our work from [10] and [11] in the following manner:

- In this work, we have considered eight attacks instead of total nine attacks included in the original UNSW dataset. We have excluded Fuzzer attack from the dataset as it is not commonly found in the multi-cloud environments [10]. Such attacks generate random packets to stop or suspend a service [10]. However, redundancy and geographical distribution are important features of services implemented in the multi-cloud environments [28]. Hence, it is hard to achieve the goals of Fuzzer attack in multi-cloud environments and easy to recover from it using the redundant service instances.
- In addition, we use a feature selection scheme to reduce the number of features while building the machine learning model. This has resulted in better performance in term of anomaly detection and prediction accuracy of anomalous traffic (presented in Section III-B). Also, we compare the RF technique against the linear regression (LR) technique to demonstrate a better performance by the RF.
- The work presented in [10] and [11] focus only on anomaly detection and overall accuracy of the machine learning models. However, our focus extends that to include feature selection and categorization of different types of attacks and categorization accuracy as well.

TABLE I. UNSW DATASET STATISTICS

| Traffic Type | Training | Training % | Testing | Testing% |
|---|---|---|---|---|
| Normal | 56000 | 31.90% | 37000 | 44.90% |
| Generic | 40000 | 22.80% | 18871 | 22.90% |
| Exploits | 33393 | 19.00% | 11132 | 13.50% |
| Fuzzers | 18184 | 10.37% | 6062 | 7.30% |
| DoS | 12264 | 7.00% | 4089 | 50.00% |
| Analysis | 2000 | 1.14% | 677 | 0.82% |
| Reconnaissance | 10491 | 5.98% | 3496 | 4.24% |
| Shellcode | 1133 | 0.64% | 378 | 0.46% |
| Backdoor | 1743 | 0.99% | 583 | 0.70% |
| Worm | 130 | 0.07% | 45 | 0.05% |

## III. FEATURE SELECTION AND ANOMALY DETECTION

In this section, we present a feature selection scheme and then measure the performance of RF and LR learning algorithms with the new set of features as far as anomaly detection is considered.

### A. Feature Selection Scheme

The original UNSW dataset included 49 features extracted from all the collected traces of the network traffic. We argue that having all these 49 features may lead to both complexity and overfitting of the machine learning models [29] [30]. Complexity may arise because learning models have to go through all the features which may consume a significant amount of time. Overfitting could be a result of fitting the model too tight to the learning dataset which may lead to higher overall error rate with the testing dataset and real-time analysis. Hence, we propose to reduce the number of features to reduce model complexity while enhancing accuracy.

We have used best-first feature selection technique [31] to reduce the number of features. In this technique, the user defines a criterion to finalize the optimal number of features. Such criterion could be: maximizing overall accuracy, minimizing the false negative or false positive rate, or minimizing the classification error for a particular attack type. In the first iteration, the algorithm chooses the best feature among all that achieves the selected criterion. This feature is saved to a set of optimal features which is initially empty. In the next iteration, each of the other features is used along with the optimal subset to build the learning models. The feature with which the model performs the best is added to the subset. The algorithm keeps appending the best feature to the subset in each iteration. It stops when the next iteration does not produce an improved result compared to the current subset or the complete set. Thus, it is guaranteed that the selected subset gives the least error in prediction and hence may be considered as an optimal subset of features.

We conducted an experiment to demonstrate the enhancement in the performance of learning models with the reduced number of features. The selection criterion was the minimum testing error using both the LR and RF learning models. Fig. 1 shows the result of the experiment where the x-axis is the number of features, and the y-axis is the relative time that learning process took. Relative time is the time required to build the model with the current number of features divided by the maximum time taken with all features. We observe that the relative time can be reduced to half if the number of features is reduced to 20.

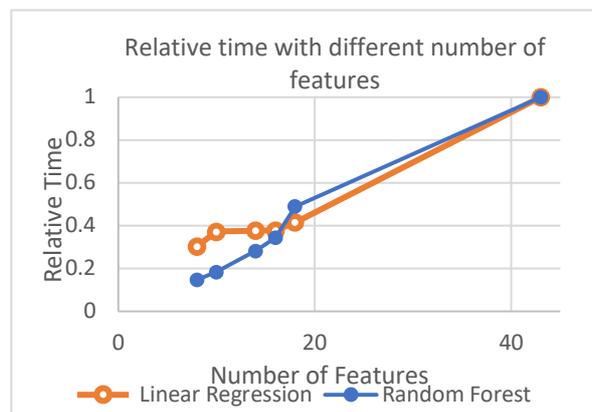

Figure 1. Number of feature versus relative time.

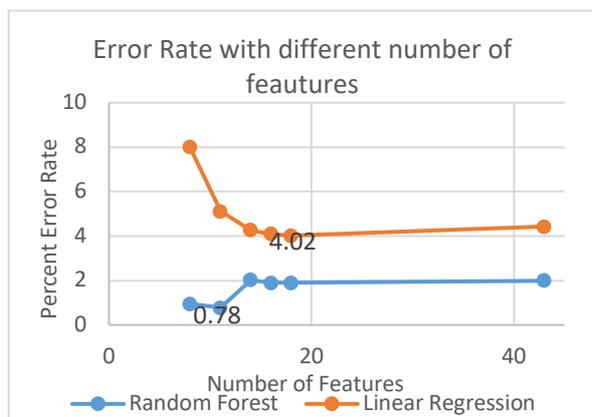

Figure 2. Number of feature versus percent error rate.

Fig. 2 shows the error rate with different numbers of features. It shows that one can reach an optimal performance with only 18 features using the LR model and 11 features using the RF model. As can be seen from the figure, increasing number of feature enhance the accuracy until it reaches an optimal performance. Then, the performance starts detroiting due to overfitting the model.

## B. Anomaly Detection using Machine Learning Models

After selecting the optimal set of features, we used supervised machine learning techniques to build the anomaly detection models. We specifically used LR [32] and RF [33] due to their simple learning models and better performance with the UNSW dataset. Both algorithms have been widely adopted in machine learning, especially for the development of the IDS [7]. SVM is another widely used algorithm. However, we did not include it in our study due to its higher complexity and lower performance than RF with the UNSW dataset.

With the experimental results obtained, we observe that the anomaly detection error rate is as low as 1% with RF scheme and 11 features. With LR, the minimum error rate that could be obtained is 4.5% with 18 optimal features. However, the overall error rate is not enough to determine the performance of any scheme, especially in the case of cloud security. False alarm rate (FAR) and un-detection rate (UND) are equally important as measures for the performance of machine learning models in the security domain. Hence, we measure FAR and UND as well to check the performance of RF and LR models. FAR is the percentage of the normal traffic which has been misclassified as anomalous by the model (Eq. 1). UND is opposite of the FAR and is the percentage of the traffic which represents an anomaly but misclassified as normal by the model (Eq. 2). The overall error is given by Eq. 3 while the accuracy is given by Eq. 4. In this paper, we aim to minimize the overall error rather than just UND or FAR.

$$FAR = \frac{false\ posative}{true\ negative + false\ posative} \times 100\% \quad (1)$$

$$UND = \frac{false\ negative}{false\ negative + true\ posative} \times 100\% \quad (2)$$

$$Overall\ Error = \frac{false\ posative + false\ negative}{number\ of\ samples} \times 100\% \quad (3)$$

$$Overall\ Accuracy = \frac{true\ posative + true\ negative}{number\ of\ samples} \times 100\% \quad (4)$$

Here, a false positive is the number of normal packets which are misidentified as anomalous packets. Similarly, true negative is the number of correctly identified normal packets; false negative is the number of anomalous packets detected as normal and true positive is the number of correctly detected anomalous packets.

As shown in Fig. 3, RF outperforms LR in all metrics. RF demonstrated an overall error of 0.9% with 0.4% UND and 1.9% FAR. On the contrary, with LR, UND is 3.5%, which is worse than that of RF. With the results presented in Fig. 3, we can safely conclude that RF performs better than LR and hence we will be using RF for any further analysis. It should be noted that LR generated optimal results with 18 features, while, RF generates optimal results with only 11 features.

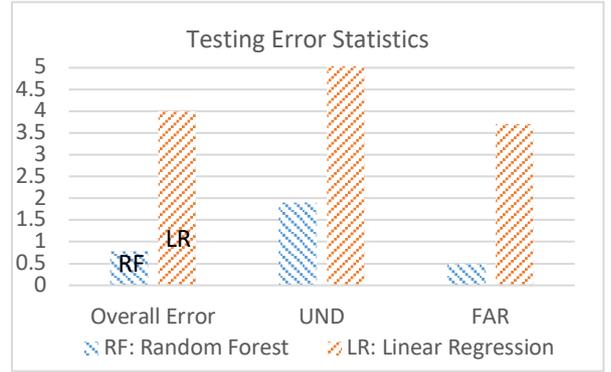

Figure 3. Anomaly detection results

## IV. CATEGORIZATION OF UNSW ATTACKS

We argue that besides determining overall anomaly detection accuracy, it is also important to distinguish different attack types individually for optimal countermeasure and defense against attacks. In this section, we first present the classical single-type attack technique used to perform attack categorization. We show the problem with such technique and then propose our own step-wise attack categorization algorithm. Furthermore, we present the results of both algorithms and argue the need for further work to reach optimal categorization results.

### A. Single-type Attack Categorization

A single-type attack model is trained with only that attack data mixed with normal traffic. For example, the DoS attack model would consider DoS attack along with normal traffic. Table II presents the results of such models without considering other attacks. It is observed that UND errors are less than 5% for Generic, DoS, Exploits, and Reconnaissance attacks using the RF learning model. This is a considerably low error rate. On the contrary, the performance is degraded for other types of attacks due to the low number of training samples for these attacks.

TABLE II: RANDOM FOREST ERROR PERCENTAGE FOR SINGLE-TYPE ATTACK DETECTION

| Type of Attack | Overall | FAR | UND |
|---|---|---|---|
| Generic | 0.22% | 0.091% | 0.39% |
| Exploits | 1.30% | 1.40% | 0.90% |
| DoS | 0.05% | 0.20% | 4.20% |
| Reconnaissance | 2.00% | 0.70% | 4.9% |
| Analysis | 0.71% | 0.56% | 7.83% |
| Backdoor | 0.11% | 0.05% | 3.70% |
| Shellcode | 0.80% | 0.60% | 11.00% |
| Worm | 0.03% | 0.00% | 20.00% |

One way of categorizing anomalies, which has been used with KDD in [21], is to use single-type attack models to separate each attack type individually. In such a way, an anomalous packet is checked against each of the single-type attack detection models till the packet is identified for a particular attack type. However, the performance of such model deteriorates as attack traffic behaviors evolve and become more correlated. Since the models are trained to distinguish between a particular attack type against the normal traffic, they result in a significant amount of misclassification errors, resulting in failure of single-type attack categorization model.

To illustrate the misclassification problem with single-type attack detection scheme using the UNSW dataset, we consider "Exploit" attack and apply its packets to different single-type attack detection models. Ideally, only "Exploit attack model" is expected to detect Exploit attacks while other models are not, as they are not trained to do so. However, when Exploit attack packets are provided to other attack detection models, such packets are observed to be detected as anomalous with DoS and Generic models, as shown in Table III. The misclassification error in Table III is the percentage of Exploit attack packets detected by the given non-Exploit attack detection models as anomalous packets. Results showcased in Table III demonstrates a high misclassification error with single-type attack detection models proposed in the literature. The misclassification error made in all types of attacks in the dataset is observed to be 20%, which is considerably high. Hence we propose a novel algorithm in the next subsection to enhance such misclassification error.

TABLE III: EXPLOITS MISCLASSIFICATION WITH SINGLE-TYPE ATTACK MODELS

| Attack Model | Miscalculation Error |
|---|---|
| Generic | 51.24% |
| DoS | 89.95% |
| Reconnaissance | 33.19% |
| Analysis | 0.00% |
| Backdoor | 15.00% |
| Shellcode | 4.00% |
| Worm | 2.00% |

*B. Step-wise Attack Categorization*

In our proposed algorithm, rather than distinguishing each attack from normal traffic, we train our models to distinguish a particular attack from other attacks traffic. First, a binary classification as normal packets against anomalous packets is performed using the RF learning model. Anomalous traffic is analyzed in further stages for different attacks categorization. In the second stage, traffic is further classified into two classes, which are Generic attacks and the rest of other attacks traffic. Note that at this stage, normal traffic is removed from the training set. Traffic with the Generic attack is detected and segregated while the rest of the traffic is forwarded to the third stage of classification. At the third stage, DoS, Exploits, Analysis and Backdoor attacks, referred to as Cat. 1 attacks, are identified as one class while the rest of the traffic is forwarded to the fourth stage. Cat.1 attacks were identified as one due to their features similarities and needed for further analysis as will be discussed next. In the fourth stage, Reconnaissance attack is identified while remaining traffic is sent to the next stage for final classification between Shellcode and Worm attacks. Note that at each stage classification is performed using the RF binary classification model, where the particular attack traffic is marked as 1 while the rest of the traffic is marked as 0, as shown in Fig.4. In this work, the order of stages was chosen based on the number of samples per attack type. Other options of ordering may be based on attack severity or classification error for a particular attack type.

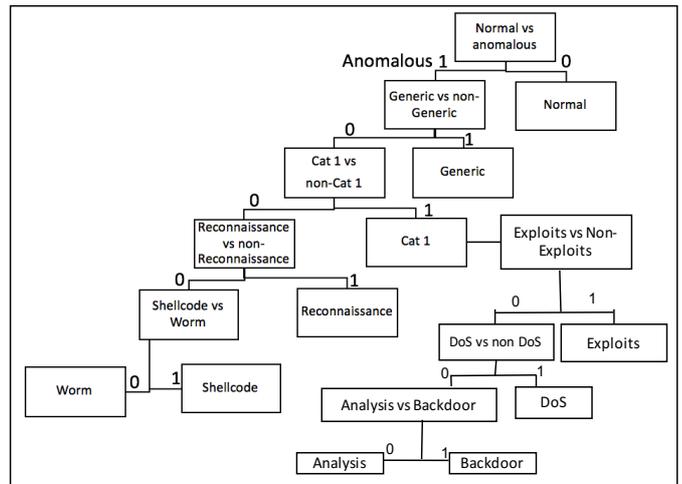

Figure 4. Proposed categorization algorithm

As can be seen DoS, Exploits, Analysis and Backdoor were merged to represent Cat. 1 attacks. Those attacks show high similarities between each other and hence they were separated to be further analyzed. We first notice that some of their features are fixed which may highly affect the accuracy. Hence, those features were removed, and only 24 features out of the 43 were considered. This enhanced the accuracy a bit, however, as will be seen later, FAR error is still very high for such models. For example, non-Exploits attacks were all detected as Exploits and hence misclassified. This can be due to biasing or high features similarities. We highlight this problem in the later analysis and argue the need for further distinguishable features or data to classify those attacks correctly.

*C. Results and Analysis*

In this subsection, we present the results of the categorization algorithm proposed in the previous subsection. In addition, we compare the results to single-type attack categorization. As indicated in the previous section, Stage 1 distinguishes anomalous packets from normal traffic and resulted in 99% accuracy. We segregate the anomalous traffic and forward it to Stage 2 which checks whether the attack is a Generic attack. We can reach around 2% UND error while only 0.06% of other attacks were misclassified in this model. Misclassification error here resulted from other attacks being classified as Generic while they are not. This is considerably good result as both errors are significantly low.

In the third stage, while classifying the traffic between combined attacks and the rest, we observe much lower accuracy. Around 20% of the Reconnaissance, Shellcode and Backdoor attacks were categorized as Cat. 1 attacks (combined attacks) in this stage. Cat.1 error is considerably low, as low as 3%. However, Exploits versus non-Exploits model results in the high misclassification as almost all DoS, Analysis and Backdoor traffics were classified as Exploits. This can be due to high biasing as Exploits had a much higher number of samples than others, demonstrated in Table I. More importantly, it can be due to high similarities between attack behaviors which make them indistinguishable unless more distinguishable features or data were added to the dataset.

At the fourth stage, the algorithm distinguishes between Reconnaissance attacks against the rest of the traffic. We observe around 0.3% UND error while around 0.5% and 13% misclassification errors were observed for Shellcode and Worm attacks at this stage. In the last stage, we distinguish between remaining two attacks, that is, Worm and Shellcode. Shellcode shows 0.3% UND error and 13% Worm attack falsely detected due to the low number of samples. Thus, the categorization error made by the proposed algorithm for all attacks is around 6.4%, which is a considerable enhancement compared to single-type of attack categorization. Worm and Shellcode have extremely low number of samples compared to others, as was shown in Table I. Thus, it is expected that their accuracy is lower than others due to the difficulty in correctly classifying them with such samples. Our algorithms showed a superior enhancement on their accuracy compared to single type categorizartion. However 70-80% accuracy is still low and can be enhanced further with the use of imbalance dataset classification as discussed in [34]. Table IV summarizes detection accuracy results (in percentage) of our proposed algorithm and compare it with single-type attack model.

TABLE IV: PERCENTAGE ACCURACY OF SINGLE-TYPE ATTACK CATEGROZATION VERSUS STEP-WISE CATEGORIZATION

| Type of Attack | Single-Type Attack Accuracy | Step-Wise Attack Accuracy |
|---|---|---|
| Normal | 99.50% | 99.50% |
| Generic | 99.00% | 97.00% |
| Exploits | 36.27% | 99.50% |
| DoS | 20.00% | 20.00% |
| Analysis | 2.00% | 2.00% |
| Backdoor | 3.00% | 5.00% |
| Reconnaissance | 40.00% | 86.00% |
| Shellcode | 20.00% | 80.00% |
| Worm | 11.00% | 70.00% |
| **Overall** | **80.00%** | **93.35%** |

Even though we observe an improvement in the overall categorization accuracy with the proposed scheme, it suffers from the following limitations. First, DoS, Analysis, and Backdoor had a low accuracy due to misclassification as Exploits in Exploits vs. Non-Exploit model. Thus, further research is needed to distinguish between these attacks by adding more distinguishable features or data or applying more sophisticated machine learning approach. Second, Reconnaissance, Shellcode, and Worm attacks accuracy need further enhancement by adding more training data or features to distinguish them from Cat.1 attacks. As was shown in Table I, the dataset samples are imbalanced and thus the use of machine learning techniques that handles such problem can enhance the accuracy in our approach. Finally, learning models that are built in this paper can be used to detect similar anomalous traffic in multi-cloud environments. However, such models need be constantly updated for new traffic patterns, modifications in the attack types and newly evolving attacks.

## V. Conclusion

In this paper, we demonstrated the feasibility of supervised machine learning techniques for anomaly detection and categorization of attacks. The UNSW dataset was chosen as it is the latest and the most comprehensive publicly available dataset. Results demonstrate that random forest (RF) technique along with feature selection scheme can achieve 99% accuracy with anomaly detection. However, we show that even if the detection accuracy is high, categorization accuracy is comparatively lower due to similar attack traffic behaviors. We used two algorithms for categorization, traditional and our own step-wise categorization algorithm. Results showed a better performance than traditional approaches. However, three of our attacks were not categorized due to feature similarities and unbalanced data. We highlight the need for more data or feature to distinguish those attacks. Finally, we argue that such models can be applied to multi-cloud environments with slight changes in the learned models.

## Acknowledgment

This paper was made possible by NPRP grant # 8-634-1-131 from the Qatar National Research Fund (a member of Qatar Foundation) and NSF grant #1547380. The statements made herein are solely the responsibility of the author [s].